# On a fundamental statistical edge principle

T. Gastaldi, Dept. Statistics and Dept. Computer Science, University of Rome "Sapienza"

tommaso.gastaldi@uniroma1.it or tommaso.gastaldi@gmail.com

**Abstract**

This paper establishes that conditioning the probability of execution of new orders on the "self-generated" **historical trading information** (HTI) of a trading **strategy** is a **necessary** condition for a **statistical trading edge**. It is shown, in particular, that, given any trading strategy ***S*** that does not use its own HTI, it is always possible to **construct** a **new** strategy ***S**** that yields a systematically increasing improvement over ***S*** in terms of profit and loss (PnL) by using the self-generated HTI. This holds true under rather general conditions that are frequently met in practice, and it is proven through a decision mechanism specifically designed to formally prove this idea. Simulations and real-world trading evidence are included for validation and illustration, respectively.

## 1. Introduction

This paper focuses on a novel result that serves to illustrate a more **general principle** that has deep implications for real-world financial applications and **rational portfolio management**, where it contributes to a deeper understanding of the actual sources of a "**systematic trading edge**" (Pardo, 2008).

To avoid losing focus on too many theoretical or application-related details, we present the concept in its most basic form. We would like to emphasize that these findings are not the result of purely abstract speculation but rather an attempt to "formalize" in the most straightforward way an observed phenomenon that initially appeared difficult to frame beyond an intuitive perception and that decades of experience in wealth management and significant advances in algorithmic trading have been increasingly pointing toward.

### A universal principle for a trading edge

We would like to substantiate the following general concept.
Let ***S*** be any strategy. At time *t*, it will have generated some **historical trading information** (**HTI**), where we assume that there is at least one or more fills so that it is not null. Denote by $HTI(t)$ such information up to time *t*.
A **sufficient** condition for any trading strategy to be inefficient and **surely dominated** by a "better" strategy is the disregard of the HTI for trading decisions. In intuitive terms, denoting by $E_t\ not\,using\ HTI(t)$ the event consisting of the execution of an order at time *t* while disregarding the current HTI and by $E_t\ using\ HTI(t)$ the event consisting of the execution of the same order while instead taking into account the current HTI, we can write, for all orders:

$$\text{Prob}(\,E_t\ not\,using\ HTI(t)\,) \;=\; \text{Prob}(\,E_t\ using\ HTI(t)\,)\ .$$

That is, if the new orders in any strategy $\boldsymbol{S}$ do not depend probabilistically on the self-generated HTI, then **there exists** a strategy $\boldsymbol{S}^*$ that instead exploits such information, which can achieve, over time, an arbitrarily large performance improvement with respect to $\boldsymbol{S}$.
This asserts the centrality of the idea of utilizing the relevant past trading information for the purpose of obtaining a **provable** statistical advantage.
Clearly, depending on how one formally defines the concept of "performance improvement", there could be as many different strategy constructions and possible proofs as there are such definitions.
In order to show this in at least **one possible** incarnation, we will prove that, given any strategy $\boldsymbol{S}$ that does not use its own HTI, there exists a **constructible** strategy $\boldsymbol{S}^*$ that improves $\boldsymbol{S}$ in the sense that:

$$\text{Prob}(\ \{ \lim_{t\to\infty}(PnL_{S^*}(t) - PnL_S(t)) = \infty \}\ ) = 1$$

where $PnL_S(t)$ is the **profit and loss** (PnL) of strategy $\boldsymbol{S}$ at time $t$, determined by the instrument price at time $t$ and all the orders filled up to time $t$. In order to accumulate such an unbounded advantage, $\boldsymbol{S}^*$ must use the self-generated HTI.
This means that, if properly exploited, the strategy's HTI can provide an objective source of (possibly additional) **statistical edge**. This result is universally true under fairly general conditions (or portfolio management), which will be discussed in greater detail in the section on assumptions.

To experimentally confirm the correctness of this specific theoretical result, we also include **forward simulations** as well as some more advanced **empirical evidence** that illustrates how the **general** concept is also fully applicable, with the required adaptations, in real-world trading activity.

## 2. Main Result

### Preliminary notation and definitions

Denote by $\boldsymbol{I}$ any **financial instrument** with a **multiplier** (or "contract multiple") $m_I$ . For simplicity, assume that the instrument does not have an expiration date or, if it does, that its price trajectory can be **rolled over perpetually** at suitable prices, possibly even switching across different financial instruments with the same multiplier and currency. Assume that the observed price trajectory is modeled as a realization of a stochastic process $\{P(t)\}_{t\in T}$, that is, a collection of random variables indexed by time, where $t \in T$ and $T$ is a set of instants in time. The price process also follows some technical assumptions, which will be discussed in a separate section, and that reflect conditions that are common in many real-world financial instruments.

Denote by $\boldsymbol{S}$ a **trading strategy**, defined as **a set of rules** (of a deterministic or non-deterministic nature) to generate, over time, **orders** in response to changes in the **price trajectory** and other possible available market data. The sequence of orders filled by $\boldsymbol{S}$ is denoted:

$$\{\ o_1, \dots, o_{n(t)}\ \},\ t \in T$$

where $t$ indicates an instant of time and $n(t)$ is the number of orders filled up to and including the time $t$. Sometimes, the set of filled orders of a strategy will also be referred to as the "**order cloud**" (referring to its visual appearance when orders are plotted).

It is also useful, for notational compactness, to introduce the following (dichotomous) function:

$$s_h \equiv 1 - (2 \cdot 1_{[O_h \text{ is an order to buy}]})$$

returning **1**, for a **sell order**, or **-1** , for a **buy order**. This could be referred to as the "**PnL-contribution sign**" because it indicates the "direction" of the contribution of the order value's contribution to the total PnL.

For the sake of simplicity, we can ignore the fact that, in the real world, there are two price trajectories for each instrument: the **bid** price (where **sell** orders are executed by traders) and **the** ask price (where **buy** orders are executed by traders), which run roughly "parallel" with a varying distance (depending on liquidity), i.e., the **bid/ask spread**. In other words, we can simply assume a spread of zero to avoid cluttering the notation with unimportant details. On the same note, **any trading expenses** (such as commissions, spreads, and the possible rollover costs) can also be ignored because, when comparing the two strategies ***S*** and ***S****, the number of orders filled and their quantities will be the same in both, and **the difference between the respective PnLs will be independent** of any trading expenses. However, for greater realism, we have included commissions and bid/ask spreads in our **simulation studies**, as well as the actual **tick sizes** and **multipliers** of real-world instruments.
With the above premises and using the sequential index $h$ as a simple order identifier, we can denote an order by the following tuple:

$$o_h \equiv (t_h, s_h, p_h, q_h)$$

where $t_h$ is the **fill time**, $s_h$ is equal to 1 for a sell order or $-1$ for a buy order, $p_h$ is the average fill **price**, and $q_h$ is the filled **quantity**. (One could also assume that a possible total order commission $c_h$ was used to adjust the price as follows: $p_h - s_h\, c_h / (m_I\, q_h)$ , but this is an unimportant detail.)

To write a convenient expression for the **PnL**, given an **arbitrary** sequence of orders up to time $t$, we can assume that a **lot matching method** is used to identify matching sell-buy quantities within the two ordered sequences of sell and buy orders (for instance, FIFO is a popular method, but in this context, the matching method is totally irrelevant as the PnL itself is **invariant** with respect to it).
Denote by:

$$M_m \equiv (p_m^{Sell}, p_m^{Buy}, q_m),\ m = 1, \ldots, m(t)$$

the lot matches, where $p_m^{Sell}$ is the **sell-side** price, $p_m^{Buy}$ is the **buy-side** price, $q_m$ is the **matched quantity**, and $m(t)$ is the total number of matches resulting from the filled orders.

Let

$$u_r \equiv (s_r, p_r, q_r),\ r = 1, \ldots, r(t)$$

be the remaining unmatched quantities (including one possible residual quantity from the last match, if any), which could result in either all buy or all sell orders, where $r(t)$ is the number of such terms.

Given any lot matching method, the PnL at time $t$ can be decomposed as follows as a result of the matching:

$$PnL_S(t) = R(t) + U(t)$$

where $R(t)$ and $U(t)$ are the "**realized**" and "**unrealized**" parts of the PnL, respectively, and $P(t)$ is the instrument price at time $t$ :

$$R(t) \equiv m_I \sum_{m=1}^{m(t)} (p_m^{Sell} - p_m^{Buy})\, q_m \,, \quad U(t) \equiv m_I \sum_{r=1}^{r(t)} s_r\,(p_r - P(t))\, q_r \,.$$

In general, these two components depend on the matching method and the current price.

On the other hand, by regarding all orders as belonging to the "unrealized" (that is, by assigning no orders in sum $R(t)$ of matched lots), the PnL can be written equivalently more synthetically as the sum, over all orders, of the **signed weighted differences** between the prices of the orders and the price $P(t)$ :

$$PnL_S(t) = m_I \sum_{h=1}^{n(t)} s_h\,(p_h - P(t))\, q_h \,.$$

For real-world trading algorithms, in which it is critical to maintain accuracy to the **tick size**, one may want to include the "closing expenses" as well, i.e., the **commissions** to close the current position and the current **spread**. For simplicity, we can disregard this detail because it **makes no difference in the proof** of our result. However, they have been incorporated into the simulation study program to increase realism.

The preceding PnL expression can be simplified further by denoting:

$$Pos(t) \equiv -\sum_{h=1}^{n(t)} s_h\, q_h = \sum_{\substack{h \le n(t),\\ Buy\ orders}} q_h - \sum_{\substack{u \le n(t),\\ Sell\ orders}} q_u$$

referred to as the "**signed open position**" at time *t,* and:

$$VPos(t) \equiv m_I\, P(t)\, Pos(t)$$

referred to as the "**signed value of the open position**" at time *t.*

Therefore, we can finally write:

$$PnL_S(t) = m_I \sum_{h=1}^{n(t)} s_h\, p_h\, q_h + VPos(t)$$

$$= m_I \sum_{\substack{u \le n(t),\\ Sell\,orders}} p_u\, q_u - m_I \sum_{\substack{h \le n(t),\\ Buy\ orders}} p_h\, q_h + VPos(t)$$

which, as one may intuitively expect, expresses the current PnL as the difference between the total absolute values sold and bought plus the current signed value of the open position that, on close, would contribute to either the sold value or the bought value, depending on the type of the closing order (sell, for a positive open position, or buy, for a negative open position).

**Theorem: Universal dominance of strategies exploiting the self-generated HTI**

Given a strategy ***S*** that does **not** use the self-generated HTI to probabilistically influence the new orders, it is possible to construct (at least) **one** more **efficient**, competing strategy ***S**** that uses its own HTI such that:

$$\text{Prob}(\ \{ \lim_{t\to\infty} (PnL_{S^*}(t) - PnL_S(t)) = \infty \}\ ) = 1\ .$$

This can be intuitively interpreted as meaning that, by exploiting the HTI of a strategy, over time one can surely amass an **arbitrarily large PnL improvement** via some suitable mechanism.

In order to construct **one** instance of such a strategy, we will use a simple "**delayed execution**" mechanism, which will allow us to easily prove the statement in a rigorous way.

**Construction of a dominant strategy that uses the self-generated HTI**

We will now proceed with the **construction** of one dominant strategy, ***S****. To enable real-time comparison, ***S**** is assumed to run simultaneously with ***S***. The strategy ***S**** is partitioned into an infinite sequence of consecutive disjoint finite **phases**, each consisting of two stages:

**Stage 1**: Strategy ***S**** runs in parallel and is identical to ***S***, executing the same orders. The duration of this stage is arbitrary but finite and must contain some fills.

**Stage 2**: The strategy employs a "**delayed execution**" mechanism, which is now described. When a new order $o_h \equiv (t_h, s_h, p_h, q_h)$ is placed by ***S***, the strategy ***S**** determines whether to execute immediately or delay an identical order $o_h^*$ according to a certain **delay event** (**DE**), defined below. If such an order is executed immediately, it is equal to the order $o_h$; otherwise, it is **delayed.** If delayed, the order is placed in a finite-length **delay queue** for later execution. The delayed order $o_h^*$ will eventually be executed exactly like the corresponding order $o_h$, with the exception of time and price, which will obviously differ. Thus, the corresponding delayed order in ***S**** is denoted as follows:

$$o_h^* \equiv (t_h^*, s_h, p_h^*, q_h)$$

where the index $h$, which is "counting" the orders in ***S***, for the corresponding orders of ***S**** is just the corresponding order identifier (and obviously does not imply a sequential ordering).

As new tickdata arrives, the strategy ***S**** continuously evaluates the possibility of executing the enqueued orders by checking a specific **execution event for delayed orders** (**EE**), which is defined below, and the enqueued orders will be executed as soon as the corresponding events become true. When there are no more delayed orders in the queue, the current phase ends, and strategy ***S**** enters a **new** phase. The expected duration of each phase is assumed to be **finite** due to the positive **recurrence** properties of the price process, at least within the considered state space, as discussed in the assumptions section below.

Let us now define the two previously mentioned events: one for **delaying** an order, and one for **executing** it.

First of all, given the sequence of orders $\{o_h\}_{h\in\{1,\ldots,n(t)\}}$ of any strategy ***S*** up to time $t$, where $\overline{p}_{Sell}(t)$ and $\overline{p}_{Buy}(t)$ are the **weighted average prices** of the sell and buy orders, respectively, up to time $t$, and $Q_{Sell}(t)$, $Q_{Buy}(t)$ are the total absolute quantities transacted in the same orders, we consider the following function:

$$C_S(t) \equiv \frac{Q_{Sell}(t)\ \overline{p}_{Sell}(t)\ +\ Q_{Buy}(t)\ \overline{p}_{Buy}(t)}{Q_{Sell}(t)\ +\ Q_{Buy}(t)}$$

referred to as the "**order cloud gravity center**" if, at time $t$, we have at least one filled order; otherwise, it can be left **undefined**. The given name reflects that it is equal to the **weighted average** of the prices of all orders up to time $t$.

**DE: Delay event**

For the strategy ***S****, define:

$$T_{S^*}(t,\ s_h) \equiv C_{S^*}(t) - s_h\ \tau$$

referred to as a "**tolerance threshold**" at time $t$, where $\tau$ is a **positive** "tolerance" amount, and also define:

$$\delta^T_{S^*,h}(t) \equiv s_h\left(P(t) - T_{S^*}(t,\ s_h)\right)$$

referred to as the "**signed distance between the current price and the tolerance threshold**" at time $t$.

Let **B(*p*)** be a Bernoulli random variable with **strictly positive** "delay probability" $p$, and, (assuming that the gravity center is not in the undefined state for the strategy ***S****) define a **"delay event"** as:

$$\{\ \mathbf{B}(p)=1\ \ \cap\ \ \delta^T_{S^*,h}(t) < 0\ \}\ .$$

When a new order is evaluated for possible execution, if the above delay event happens to be false or the gravity center $C_{S^*}(t)$ is still undefined, the order is **normally executed at the** **current price**; otherwise, the order is delayed and placed into an execution **queue** for later execution. The maximum number of delayed orders in the queue (e.g., one, two, or more) is **finite** (it can be a user parameter). If the imposed length **limit** is reached, any new order is executed normally without delay.

[For practical purposes, as more meaningful variants that avoid enqueuing orders that are too close together in terms of price, some conditions dictating a minimum distance between enqueued orders could be introduced, e.g.:

$$s_h\ (\mathit{price\ of\ the\ enqueued\ order} - \mathit{price\ of\ the\ new\ candidate\ order}\ ) > \mathit{min\ distance}$$

for all enqueued orders, where *min distance* denotes some user-defined function. The modalities for implementing user-defined filters on the enqueued orders that meet the condition $\delta^T_{S^*,h}(t) < 0$ are obviously limitless. For generality, these possibilities may be collectively represented by using the random variable **B**, as any arbitrary changes in the choice of which specific orders, among all possible candidates, are being enqueued.]

**EE: Execution event**

Let's now define the "**execution event**" for delayed orders.

It is convenient, for notational compactness, to define the following function:

$$\mathit{minmax}_s(x, y) \equiv \frac{x+y}{2} + s\ \frac{x-y}{2}$$

as a simple way to return either the minimum or the maximum of $x$ and $y,$ depending on a sign variable $s$ (the *min*, for $s = -1$, or the *max*, for $s = 1$), since the semi-distance of the two values is being subtracted or added from their midpoint.

Recall that $t_h$ and $t_h^*$ denote the instants when the order $o_h^*$ is delayed and filled, respectively. Analogous to the previous case, for any enqueued order waiting to be executed, consider the following threshold level at time $t$ :

$$\begin{aligned} G_{S^*}(t, t_h, s_h) &\equiv \mathit{minmax}_{s_h}(C_{S^*}(t) + s_h\,\gamma,\ C_{S^*}(t_h) + s_h\,\gamma) \\ &= \mathit{minmax}_{s_h}(C_{S^*}(t),\ C_{S^*}(t_h)) + s_h\,\gamma \end{aligned}$$

referred to as a **"gain threshold"**, where $\gamma$ is a **positive** "gain" amount (a user parameter).

Define:

$$\delta^G_{S^*,h}(t) \equiv s_h\,(P(t) - G_{S^*}(t, t_h, s_h))$$

referred to as the "**signed distance between the current price and the gain threshold**" at time *t*.

The **"execution event"** for delayed orders is then defined as:

$$\{ \ \delta^{G}_{S^*,h}(t) > 0 \ \} .$$

The reason for keeping track of both values of the gravity centers, the current one $C_{S^*}(t)$ and that at delay time $C_{S^*}(t_h)$ , in this execution event is that, even if the current price $P(t)$ exceeds a gain threshold based on the previous value $C_{S^*}(t_h)$ , by the same criterion, intuitively it would not still be suitable for execution according to the **updated** gain threshold at time $t$ based on the current value $C_{S^*}(t)$ . Therefore, for conceptual consistency, for the gain threshold at time $P(t)$ we consider both the **present** and **past** vertical location (or "*weighted altitude*") of the order cloud.

The above definitions can be intuitively regarded as a way of "summarizing" and "packing" into **functions of the trading data** the **HTI** accumulated by the strategy ***S****. In fact, the gravity center $C_{S^*}(t)$ of the order cloud, which changes dynamically over time, in this case serves as a moving "**structural reference line**" that incorporates past trading information on how **future orders can be influenced probabilistically** to generate better trading decisions. The use of such information from an intuitive standpoint represents an effort to create a form of **"adaptive self-consistency" of the order cloud** to favor a more profitable "development" over time of its overall shape.
The HTI exploitation also takes advantage of the possible recurrence properties of the price process or, for suitable financial instruments (e.g., rapidly decaying), of the price "revisitation" **induced** by the folio manager through perpetual **rollovers**. With such structural information, even the most basic mechanism for improving trading decisions can allow us to reduce the probability of orders that negatively impact the global PnL.

The delayed execution mechanism is only **one** simple example, among **infinite** possibilities, of exploitation of the HTI, whereas the preceding theorem points to a deeper and more general intuitive meaning. Indeed, it indicates that there is a **necessary** source of probabilistic **edge** that is frequently overlooked, with the exception of a small group of more intuitive and insightful financial actors. This source is **distinct** from the possible **information coming from market data**, which **rational** portfolio managers who subscribe to the efficient market hypothesis (**EMH**) may even dismiss as providing **no strategic advantage**.
Concerning the EMH, much evidence about it was accumulated mostly in the 1960s and 1970s, to which the proponents of "behavioral finance" responded by pointing out possible mechanisms of violations or "anomalies", such as, for instance, short-horizon reversal, medium-horizon momentum, long-horizon overreaction, as well as the persistence of volatility (cf. also the "adaptive market hypothesis"). However, as discussed, for instance, in Fabozzi and Shirvani (2020), some of these may not be "real" phenomena and have alternative explanations within the **rational expectations theory**, under "proper distributional assumptions for the historical returns".

[Some details now follow, which can be skipped on first reading, as they are mostly useful for implementation.]

**Detail a)**
The positive threshold quantities $\gamma$ and $\tau$ are **not** meant to necessarily be absolute constants. On the contrary, in general, they can be arbitrary **functions of the price process**. For instance, they could be specified as a percentage of either some reference price or of the **current price**.

They could also be defined as functions of the instrument's **volatility** or the current "*height*" of the order cloud (the difference between the maximum and minimum price of orders). This would also result in an infinite number of implementation variations, all of which are formally included in our treatment.

**Detail b)**
The **delay event** and the **execution event** can also be written in the following equivalent forms, more useful for implementation purposes:

$$
\begin{aligned}
\delta^{T}_{S^*,h}(t) < 0 &\Leftrightarrow s_h\ (P(t) - T_{S^*}(t, s_h)) < 0 \\
&\Leftrightarrow s_h\ (P(t) - C_{S^*}(t) + s_h\ \tau) < 0 \\
&\Leftrightarrow s_h\ (C_{S^*}(t) - P(t)) > \tau
\end{aligned}
$$

$$
\begin{aligned}
\delta^{G}_{S^*,h}(t) > 0 &\Leftrightarrow s_h (P(t) - G_{S^*}(t, t_h, s_h)) > 0 \\
&\Leftrightarrow s_h (P(t) - minmax_{s_h}(C_{S^*}(t), C_{S^*}(t_h)) - s_h\ \gamma) > 0 \\
&\Leftrightarrow s_h (P(t) - minmax_{s_h}(C_{S^*}(t), C_{S^*}(t_h))\ ) > \gamma
\end{aligned}
$$

**Detail c)**
The presence of the Bernoulli trial in the delayed event definition is not strictly necessary. The reason for inclusion is that it represents a way to make the formal treatment more abstract by signifying that one could also define any sort of arbitrary "filter" for the orders being delayed (thus creating infinite implementation variants) without damaging the formal conclusions. For example, if some specific orders in a real-world strategy cannot be delayed for whatever reason (for instance, because they are considered significant for **hedging** purposes), it does not matter in the grand scheme of things, as we are dealing with asymptotic behavior. Other ideas could include applying appropriate filters to the enqueued orders to possibly reduce the expected recurrence time, and so on.

## 3. Forward simulations and anecdotal empirical evidence

For illustrative purposes, we have gathered some "companion material" to this article [including several **forward simulations**, source code, and some **empirical evidence of actual trading** on real accounts] on a web page at the following permanent link:

**https://www.datatime.eu/public/arXiv_paper/**

[The page also refers to **real trading experiments,** providing examples of live trading results obtained over several years on large accounts. This also demonstrates how it is possible to automate an effective "money-printing machine" based on "structural" decay and "induced" price "recurrence". Of course, in actual implementation, the matter becomes very complex, and it is critical to take a well-designed **continuous** automated **hedging** action and to constantly monitor the **margin requirements** in order to avoid being forced to take losses at any stage, which would obviously slow down the profit accumulation process and postpone profits.
Clearly, the real trading evidence presented is purely **anecdotal**, and it is not intended to "prove" anything, as the results in the real world are obviously also influenced by a variety of factors such as software implementation, actual on-field experience, and, most importantly, optimal risk-management rules. The author can provide more real-money algorithmic trading cases that strictly adhere to the principle explained here. Refer to the corresponding section of this paper for a conceptual "proof".]

In the simulation study, we use simulated instruments whose specifications are the same as the corresponding real-world instruments. **Drift** or **volatility** parameters can also be supplied in the specifications, along with **tick size**, **tick rate**, and **multiplier**. We make sure that the **bid** and **ask** prices are all **exactly** on the corresponding **discrete price grid**. In these simulations, we use a quadruple-precision, 128-bit decimal floating-point format type for quotes, as appropriate for financial applications where exact decimal rounding is necessary.

These simulations use **general random walks**, with **actual tick sizes** and **random orders**, in order to simplify the model. This scheme is obviously **not** intended to be "profitable" in any way because the orders are simply discretized Poisson uniform arrivals (at a user-definable rate and with an additional constraint on the minimum distance between fills), and it is obvious that from this purely random market **cannot** originate any **information** useful for an edge. In practice, this is conceptually equivalent to implementing a weak form of **efficient market hypothesis** (**EMH**), where the possible **systematically increasing PnL difference** between ***S**** and ***S*** that we observe can be attributed solely to the exploitation of the **HTI**.

One example of simulation is shown in the image below, which displays the PnL curves of ***S*** and ***S**** (bottom-right chart) and the random trajectories of the bid and ask prices with random orders (top-right chart). It is also visually evident how ***S**** (red line) systematically improves ***S*** (blue line) over time at the end of each phase (vertical gray lines):

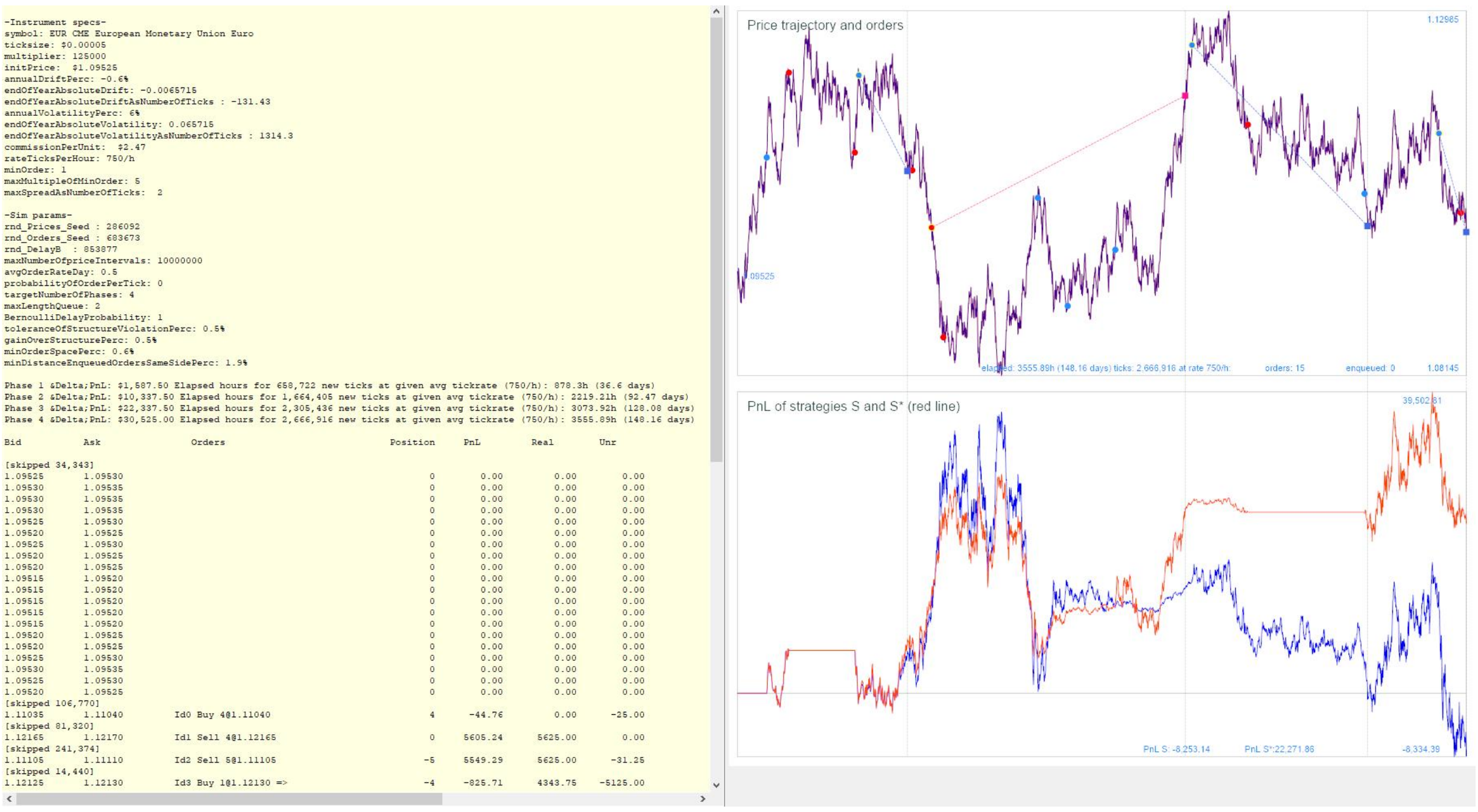

***Figure 1***. *A random example of a forward simulation (the full set of simulations, real-world examples, and source code can be found on the linked website).*

The computational results of the simulation study confirm the theoretical result of the above theorem: **at the end of each new phase**, the difference in PnL between ***S**** and ***S*** is strictly positive and increases monotonically.

The random processes (random walks with real-world tick sizes) used in these simulations only locally resemble the price curves of real instruments in the relatively short term. It is worth noting that, in theory, a pure random walk would **not** even satisfy some of our theoretic assumptions (discussed later) because it is **null recurrent** (Pólya, 1921) and has volatility proportional to the square root of time, making it obviously totally unsuitable for any financial instrument's long-term model. Similar considerations could be made for the most common theoretical **recurrent** processes, including, for instance, martingales (essentially, the idea that the average of the future price is equal to the current price) with infinite quadratic variation (Durret, 2019). Undoubtedly, more realistic and complex models could yield larger PnL improvements. However, the goal of these simulations is **not** to support a specific idea for gaining dominance under a particular price model but only to show, with the least effort, the correctness of the formal result used to illustrate the more general principle (which is largely insensitive to deviations from probabilistic assumptions).

In real-world applications, obviously, we also had to take into account the specific characteristics of traded securities, and the absolute necessity of strictly controlling margin usage, tail **risk**, and its actual manifestation in unavoidable **drawdowns** became critical. As a result, it has been necessary to implement full automation and introduce very reactive forms of **dynamic hedging** along with "**stop-loss-order recovery mechanisms**", as cursorily mentioned below. In academic financial models, it has become commonplace, for relative ease of statistical treatment, to use **variance** or also volatility (the degree of variation of a trading price series over time, e.g., the standard deviation of the log return) as a method of evaluating the **risk** of an investment. However, for our application, we prefer to focus more properly on downside expressions of risk because that is what is **actually perceived and understood by real investors**, while when volatility works in our favor, obviously it can only be welcomed. For evident reason, in this paper we cannot indulge in explanations about our real-world applications and the innumerable details and complications of the software implementation of reliable real-time trading automation because, in any case, whatever "empirical evidence" is presented, in this context, it would be considered of merely "**anecdotal**" value, certainly useful to corroborate the general idea, but ultimately we always need to rely only on formal "proofs" for the general principle.

The following picture shows a typical shape and drift in the PNL curve.

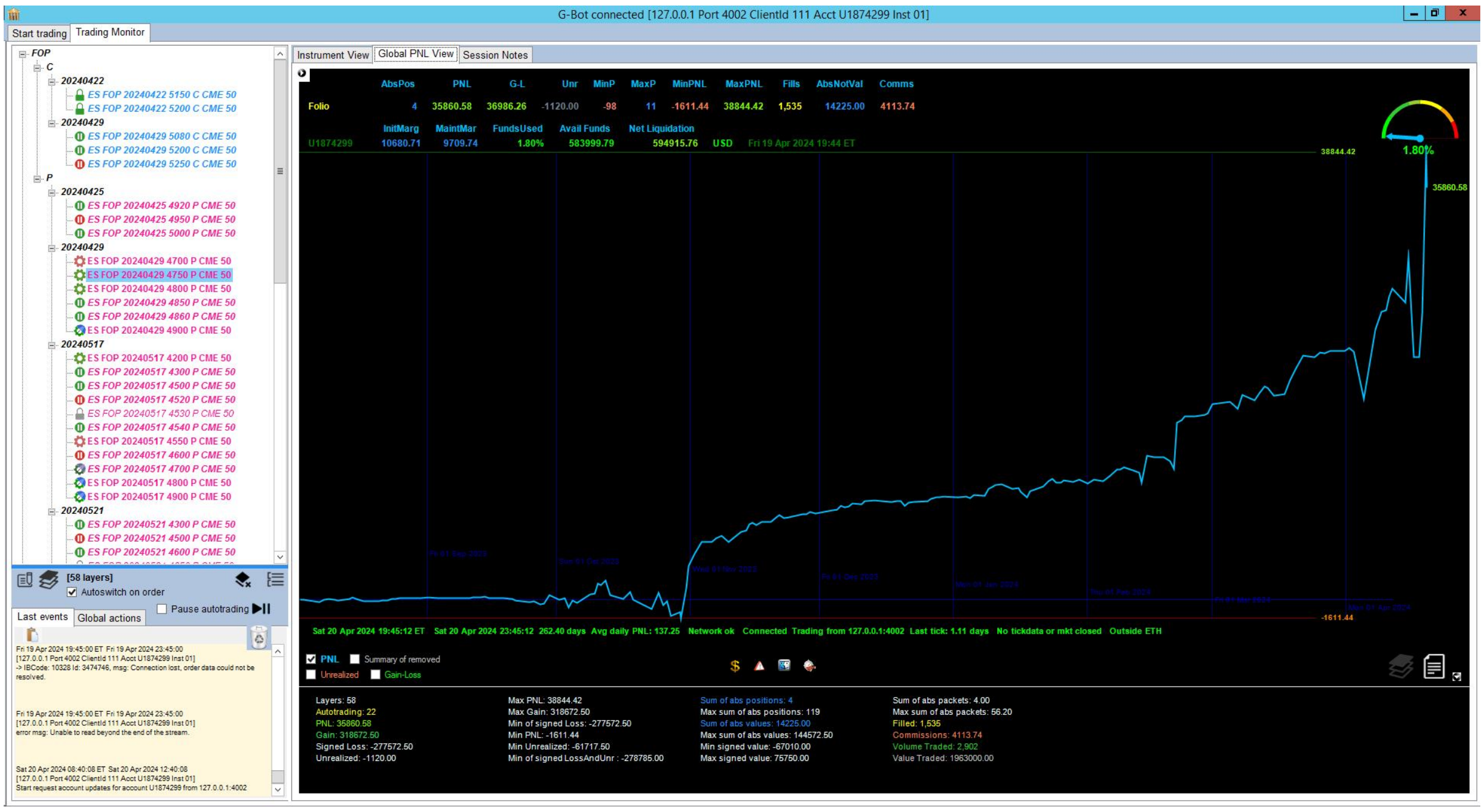

***Figure 2**. Typical shape and drift exhibit by the PNL curve in real-money applications. (More examples, with larger initial capital, and videos can be found following the provided link.)*

## 4. Philosophical and intuitive aspects

Our goal was to demonstrate, in the simplest possible way, the general principle that **any** strategy that does **not** use the self-generated **HTI** can, in principle, be **dominated** by suitably exploiting the HTI. For instance, the delayed execution mechanism that we have used essentially makes use of the past trading information along with the **recurrence** properties of the price wave to "improve" some trading decisions in the future.
Price recurrence can occur due to the **intrinsic randomness** of the process or, in practice, it can be "**induced**" through the perpetual **rollover** of suitably chosen financial instruments (as an example, you might think of OTM options).

The subject of HTI exploitation also provides suggestions for resolving the more philosophical consideration of what constitutes (in a purely conceptual sense) a "**realized** loss". It could be argued that once the **information** about a loss is "forgotten" by the strategy (that is, the relevant HTI has **no influence on the probability** of any new order), it can no longer contribute to performance improvement. In this (purely conceptual) sense, "realizing a loss" has little to do with moving it from the **unrealized** to the **realized** component of the PnL (aside from **accounting terminology** where the **realized** and **unrealized** have a well-defined meaning for capital gain taxation). While at the accounting level some "losses" can be arbitrarily switched between "realized" and "unrealized" by opening or closing out positions and lot matching, from a purely conceptual standpoint, it is instead the disregard for the corresponding relevant **trading information** (i.e., "**forgetting the losses**") for future trading activity that makes them (conceptually) "realized". In fact, if they are taken into account, it is possible to exploit the HTI incorporated in those "losses" to contribute to a **statistical edge**, both in **principle** and in **practice**, by suitably combining the relevant information with price dynamics and folio management. Similar considerations can be made for "realized" gains.

It should be noted that most trading approaches currently **ignore** the HTI, particularly in the algorithmic field. The majority of algorithmic trading strategies (excluding market-making algorithms, obviously) that we see struggling with **price unpredictability** are essentially placing orders without exploiting past trading information and relying solely on key **technical indicators** (based on price, volume, open interest, and so on). By doing so, they are implicitly presuming that there must be something to "**learn**" from the market data that would create some form of edge. Similarly, when using "**machine learning**" techniques to submit orders, one is implicitly **taking for granted** that **market data has to contain exploitable information** that can be harvested to generate a **systematic advantage**. These are, however, mere arbitrary suppositions based on **unfalsifiable** beliefs, often nourished by the misinterpretation of **backtests**, where never-failing **curve fitting** feeds fallacious expectations. Other times, especially in the high-frequency space, it is the ubiquitous **look-ahead bias** that creates, in the simulations, an insidious and dangerous illusion of profitability. This often occurs when the filled orders are so relatively close together, compared to the timeframe used for the indicators, that a local interpolation is easily exploited to generate profits that are never to be seen in the real world.

In general terms, the central message to carry away is that, by applying probabilistic conditioning of the orders on the HTI, it is always possible to construct a new strategy that, in probabilistic terms, improves a strategy that does not use it. In this sense, one could say that any strategy ***S*** that does not use the HTI is **inadmissible** because it is always **dominated** by a strategy ***S**** (and therefore infinite variations of it) that **can be actually constructed**.

This is, in our view, a **foundational** concept. Indeed, the fact that a formal argument can be provided for the HTI being a **necessary** source of objective edge should immediately raise the priority of including it among the possible sources of edge. Especially when compared with otherwise **unfalsifiable** ideas opposing the **EMH**, where different experts may often have divergent "opinions" about whether it is possible to extract information from the market data useful to gain an edge, essentially creating mere philosophical stances that ultimately cannot be proven to provide any systematic edge for real-world trading performances.

## 5. Proof of the theorem

Consider the time $t$ at the end of any phase of strategy ***S****. At this time, by construction, the two strategies ***S*** and ***S**** shall have filled the same orders, except that some of them in ***S****, namely the delayed orders $o_h^*$, will have a different time and price than the corresponding orders $o_h$ in ***S***. Thus, if we evaluate the difference of the PnLs of the two strategies, ***S*** and ***S****, and consider the corresponding orders, the non-delayed orders (and **any** trading expenses, including those of possible rollovers) **cancel out** because they are equal and opposite. As a result, the terms that remain are only those involving delayed orders:

$$PnL_{S^*}(t) - PnL_S(t) =$$

$$= m_I \sum_{\substack{h \leq n(t), \\ Delayed\ orders}} s_h\, p_h^*\, q_h + VPos(t) - \Big( m_I \sum_{\substack{u \leq n(t), \\ Delayed\ orders}} s_u\, p_u\, q_u + VPos(t) \Big)$$

where the terms relating to the current position values and their closing costs cancel out, and thus, we get the sum of the signed differences of prices between all the delayed orders in ***S**** and the corresponding orders in ***S*** :

$$= m_I \sum_{\substack{h \le n(t), \\ Delayed\ orders}} s_h \, (p_h^* - p_h) \, q_h$$

by adding and subtracting the thresholds $T_{S^*}(t_h, s_h)$ [assumed to be inside a suitable subset of prices $\mathbb{E}_s$ , as defined below in the assumptions section] and the order cloud gravity center $C_{S^*}(t_h)$ at time $t_h$ , we get:

$$= m_I \sum_{\substack{h \le n(t), \\ Delayed\ orders}} s_h \; ( \, (p_h^* - G_{S^*}(t_h^*, t_h, s_h)) + (G_{S^*}(t_h^*, t_h, s_h) - C_{S^*}(t_h))$$

$$+ (C_{S^*}(t_h) - T_{S^*}(t_h, s_h)) + (T_{S^*}(t_h, s_h) - p_h) \, ) \; q_h$$

recalling that $t_h$ also represents the time at the delay event, and that $t_h^*$ is the time at the execution event for the order $o_h^*$ , and, since the quantity $s_h \, (G_{S^*}(t_h^*, t_h, s_h) - C_{S^*}(t_h))$ is greater than or equal to $s_h \, (G_{S^*}(t_h^*, t_h, s_h) - minmax_{s_h}(C_{S^*}(t_h^*), C_{S^*}(t_h)))$ , the following inequality follows:

$$\ge m_I \sum_{\substack{h \le n(t), \\ Delayed\ orders}} s_h \; ( \, (p_h^* - G_{S^*}(t_h^*, t_h, s_h)) + (G_{S^*}(t_h^*, t_h, s_h) - minmax_{s_h}(C_{S^*}(t_h^*), C_{S^*}(t_h))\,)$$

$$+ (C_{S^*}(t_h) - T_{S^*}(t_h, s_h)) + (T_{S^*}(t_h, s_h) - p_h) \, ) \; q_h$$

from which, by the definitions of the two thresholds $\gamma$, $\tau$ and assuming for simplicity that they remain constant over time, we get:

$$= m_I \sum_{\substack{h \le n(t), \\ Delayed\ orders}} s_h \; ( \, (p_h^* - G_{S^*}(t_h^*, t_h, s_h)) + s_h \gamma + s_h \tau + (T_{S^*}(t_h, s_h) - p_h) \, ) \; q_h$$

and, since the prices at delay and execution time are, for each order, respectively $p_h = P(t_h)$ , and $p_h^* = P(t_h^*)$ , we can write:

$$= m_I \sum_{\substack{h \le n(t), \\ Delayed\ orders}} s_h \, ( \, (P(t_h^*) - G_{S^*}(t_h^*, t_h, s_h)) - (P(t_h) - T_{S^*}(t_h, s_h)) + s_h \gamma + s_h \tau \, ) \; q_h$$

$$= m_I \; ( \sum_{\substack{h \le n(t), \\ Delayed\ orders}} (\delta^G_{S^*,h}(t_h^*) - \delta^T_{S^*,h}(t_h)) \, q_h + (\gamma + \tau) \sum_{\substack{u \le n(t), \\ Delayed\ orders}} s_u^2 \, q_u \; )$$

($s_h^2 = 1$ because, by definition, it is the square of either 1 or -1) therefore, denoting by $Q_D(t)$ the total absolute quantity transacted in delayed orders up to time $t$, we finally obtain:

$$= m_I\, Q_D(t)(\gamma + \tau) \; + \; m_I \sum_{\substack{h \leq n(t), \\ Delayed\ orders}} (\delta^G_{S^*,h}(t^*_h) - \delta^T_{S^*,h}(t_h))\, q_h \; > 0$$

where each term in the above summation is positive because, for each order executed and filled at time $t^*_h$, we have $\delta^G_{S^*,h}(t^*_h) > 0$ by the **execution event** definition**,** while, since the same order was earlier **delayed** at time $t_h$, we have $\delta^T_{S^*,h}(t_h) < 0$ by the **delay event** definition (in case the thresholds are not taken constant with respect to time, they obviously remain inside the summation while the inequality still holds)**.**

This means that, at the end of each phase, the PnL difference between ***S**** and ***S*** is strictly **positive** and **monotonically increasing** because the new orders of each phase will add new **strictly positive terms** to the above summation. As a result, given a discrete state space with finitely many possible prices, as it is actually the case with real-world financial instruments (and as we formally assume for the underlying stochastic process), the PnL difference diverges to positive infinity with probability 1.

## 6. Assumptions

It should be intuitive to visualize why the stated theorem can apply to several real-world instruments, based on everyday experience with continuous price fluctuations in **STKs**, **ETFs**, **OPTs**, **FUTs**, **FOPs**, etc. However, since we make a statement of a probabilistic nature about **strategic dominance,** we also need to address the formal conditions for the general result to theoretically hold.

**Assumptions on the strategy**
Since the statement is asymptotic, we first need to assume that a suitable trading strategy is carried out continuously, while a stochastic process realizes its price trajectory over time.

**Assumptions on the price process**
Regarding the price process, we would like to make the fewest and most down-to-earth assumptions possible. In particular, we do not need to involve unrealistic idealized theoretical models like continuous-space-time stochastic processes. In fact, in the real world, prices can **only** move by a positive decimal **tick size** (cf. **decimalization** process), that is, the smallest price amount a security can move in an exchange, and only take values within a **finite and discrete** set of prices. (Examples of tick size are: 0.01 for most stocks, or 0.25 for the S&P futures ES, 0.01 for crude oil futures CL, 0.005 for silver futures SI, 0.10 for gold futures GC, 5E-05 for EUR futures, and so on.) Therefore, real-world bid and ask prices can only be **multiples of the tick size** of the instrument. In fact, for instance, Baldacci et al. (2020) note: "*in actual financial markets, transaction prices are obviously lying on the discrete tick grid. This discreteness of prices is a key feature which cannot be neglected at the high frequency scale since it plays a fundamental role in the design of market making strategies in practice*".
Similarly, in the real world, quotes or transactions exist only with a timestamp attached (entries on the consolidated tape), so we can assume that time is discrete too. Therefore, for the purposes of this work, we may assume **discrete and finite space** (the price tick grid) and **discrete time (**while we will welcome extensions to different types of assumptions from further research).
As to the dynamics of the **price process**, denote by $\mathbb{E}_s$ a suitable **subset** of prices within the finite tick grid, where a strategy is assumed to submit and execute orders. From the current price, we will need to be able to exceed some top or bottom threshold price levels in $\mathbb{E}_s$ in a finite average time. This is a fairly **reasonable** form of "recurrence" for many real-world financial instruments, where price levels are clearly seen repeatedly over time. Therefore, we will require the following "above" and "below" threshold" positive recurrence properties:

$$Prob(\{\ \exists\ 0 < u < \infty \text{ such that } (P(t+u) > p + \xi_t\ )\ \}\ |\ P(t) = p\ ) = 1$$

$$Prob(\{\ \exists\ 0 < v < \infty \text{ such that } (P(t+v) < p - \varsigma_t\ )\ \}\ |\ P(t) = p\ ) = 1$$

for any positive arbitrary prices, times, and thresholds $\xi_t$, $\varsigma_t$ such that all the involved prices are within $\mathbb{E}_s$.

Apart from theoretical speculation, these assumptions do not contradict empirical experience, in which the fluctuation and **recurrence of prices is a commonly observed phenomenon** across a wide range of financial instruments and actually the most evident aspect of many price dynamics of assets. The very existence of concepts like "support" and "resistance", whether "illusory" or not, also reflects this observation, and their occurrence is sometimes justified through the antagonistic interplay between supply and demand. Forms of price recurrence are probably one of the only apparent characteristics that many financial instruments exhibit and also it may not be a coincidence that it is implied in the self-symmetry to which much the work of Mandelbrot (1963, 2006) on fractal geometry is pointing to. For example, in many **commodity futures** (e.g., energy, metals, agriculture), the recurrence of price levels beyond dynamic thresholds is not unrealistic, at least **within some suitable subset of the range of prices** to which one can restrict and center the action of a strategy.
Where, instead, the assumptions are not appropriate, **specific variants and folio management adaptations can also be developed**. For instance, instruments with a strong **drift** could **skip** the delayed execution of orders on either the buy or the sell side.

Another important way through which non-random forms of price "recurrence" can always be systematically "induced" in some suitable financial instruments is through the use of **rollovers**, as described below.

**Induced price "recurrence" obtained through rollovers**
When the required assumptions are not plausible, there is also another mechanism, besides the **intrinsic randomness** of the price, that, in some classes of instruments, could generate forms of price "recurrence" (using quotes here or using the word "**revisitation**" to differentiate it from the purely "random" recurrence) that are anyway directly exploitable by the **folio manager**. To visualize this with a concrete practical example, take, for instance, the **deep OTM FOPs** (out-of-the-money futures options) of any underlying (e.g., the S&P), which have a strong **downward drift** (decay) because they have no intrinsic value and trade on their time value. In this case, continuous rolling over suitable contracts on a regular basis (before contract expiry) in the options matrix to new suitable prices can produce a less aleatoric form of price "recurrence".

Furthermore, in this case, **new**, more powerful mechanisms for HTI exploitation could be developed. As an example, suppose one has a **short** position on a traded FOP and that there is a sudden increase in the option price. In this case, a buy order placed by a strategy may have the purpose of temporarily "stopping" the increasing negative unrealized by "packing" it into the realized component of the PnL. In many cases, such an order could be placed out of the necessity of **hedging**, and real-world applications may not be advised, or even be able, to delay some of these fills due to margin requirements, fund availability, or drawdown (DD) containment constraints. Therefore, the delayed execution mechanism could be replaced with actions of conceptually different nature that ultimately result in a similar effect and the creation of an actual edge. For instance, one could let the buy order execute to meet hedging needs and, only later, when made possible by a suitable **roll-up**, execute a sell at a higher price. Clearly, in these cases, we cannot formally talk of random "recurrence", as commonly intended in stochastic processes' theory (cf. Feller, 1949; Lampert, 1960), but, in practice, a de facto "revisitation" of prices can be directly induced at the folio management level by **rolling up** the options within the **options matrix** to a **farther away expiration date** or a **closer strike**, so that the **new price curve** reaches a level that allows "matching" the stop order with a correspondent higher sell. This yields, in practice, a sort of "**stop-loss-order recovery mechanism**" whose final effect on edge buildup conceptually is not very dissimilar from the delayed execution mechanism we have been previously theorizing, as, in order to do all that, we still need to use the **HTI**.

Indeed, when dealing with real-world financial instruments, some portfolio managers and investors may find it preferable to choose securities that deliberately violate the random recurrence assumptions and employ more manageable mechanisms for HTI exploitation. For example, by relying on the combined effects of decay and rollover to take advantage of a systematic and plannable "periodicity" (e.g., the expiration date of traded instruments), rather than relying on the unpredictable randomness of price processes, which may result in larger drawdowns of exhausting duration.

The preceding general indications should be regarded as mere cursory hints among the infinite possibilities that may emerge in real-world applications, which, as we have experienced firsthand, can rapidly grow very complex and necessitate extremely careful algorithmic implementation, where even the smallest detail can make a significant difference.

For anecdotal evidence about trading with the use of HTI in the **real world**, see also, for instance, the video example linked on the web page with the companion results to this paper (still running in real-time, and the latest update videos and official broker report can be provided anytime on request to interested readers).

## 7. Applications in the real world and future lines of research

The primary goal of this paper was to propose and demonstrate, both in theory and practice, that a **rational** search for a true **edge** cannot ignore the use of a strategy's **self-generated HTI** because its use to probabilistically improve order decisions is a **necessary** condition for the **admissibility** of any trading strategy.

In more intuitive and "visual" terms, this means that any strategy, by "looking backward" at its **trading information**, is potentially able to favorably **influence the probability of new orders** so that its **"order cloud"** can develop in the future while remaining "**self-consistent**" with its **historic evolution**, thus allowing the possibility of building a long-term edge.

Although this may even appear obvious in hindsight, not everything that seems intuitively plausible can always find a rigorous justification. For instance, the use of **technical analysis** in the algorithmic field implies that some financial actors find "intuitive" (surprisingly for us) the notion that **market data** can allow them to profit from historical price statistics descriptive of the price curve. However, this presumption comes indeed with no proof because, like its conceptual counterpart, the **EMH**, it is not **falsifiable**, and as W. Pauli would put it, "*it is not even wrong*". Finally, it should be noted that the majority of current algorithmic trading (excluding obviously market making activity) activity appears to be completely oblivious, or not explicitly aware, of the **necessity** of exploiting the **HTI** for long-term profitability.

Note that we do **not** claim any suitability of the specific elementary "construction" considered here, of a dominant strategy, for **real-world** applications. Actually, we claim the opposite, as it should be considered only as a purely **formal**, **simple** conceptual device (one of infinite possible variants), a "**thought experiment**", created on purpose to rigorously prove one instance of a more **general principle**.
Furthermore, observe that the ideas discussed here obviously do **not** necessarily imply, in principle, that one could start with a massively "unprofitable" strategy and turn it into a "profitable" one by using the HTI. In fact, the PnL difference between ***S*** and ***S**** could even grow in the presence of a systematic **accumulation of losses** in ***S*** at such a high rate that it may even prevent the PnL of ***S**** from ever getting positive if a specific effort is made in this direction. Clearly, in actual practice, such a strategy, massively accumulating losses, would have to be designed "on purpose". This is due to the fact that when trading solely on market information, the long-term average PnL is likely to be close to zero, as if trading randomly, with a negative drift due to commissions and spread. In this case, the use of HTI can make a significant difference in the PnL curve.

Furthermore, it should be noted in general that no edge or probabilistic advantage can ever absolutely "guarantee" systematic profitability in the real world. "A trading edge is a technique, observation, or approach that creates a cash advantage over other market players" (from Investopedia), but, in actual practice, even with techniques theoretically suitable to obtain a profit, one could still fail to realize it due to other reasons. For instance, a common reason is excessive fund allocation relative to the total available funds, which may cause forced losses due to the liquidation of positions in situations of marked drawdown or rapidly raising of margin requirements (where applicable) and therefore impede the practical possibility for the edge to emerge in time and manifest itself in actual profits.

In real-world applications, the quantitative analysts ("quants") are called upon to work on more realistic and pragmatic schemes in which the exploitation of the HTI should aim to meet the expectations of investors while adhering to appropriate drawdown constraints. This could take into account several features that financial instruments have, such as, for instance, decay, drift, and rollovers as in the case of FOPs (future options), daily rebalancing and reverse split as in the case of leveraged and inverse ETFs (exchange-traded funds), contango, backwardation, reversion, cyclicity, seasonality, etc., as for some commodities FUTs and STKs (futures and stocks), and so on. Therefore, we encourage the researchers to explore suitable frameworks to further extend these concepts. For instance, Ornstein-Uhlenbeck (1930), Vasicek (1977), Cox-Ingersoll-Ross (1985), and many other scholars have extensively studied price processes that **tend to return to certain levels** (such as EMR, GMR, or their variants). In these cases, appropriate probabilistic conditions could be used to control the frequency of the delay events for buy or sell orders according to the distances from suitable price levels. For **drifting** instruments, when the drift manifests itself over relatively long timeframes, as in the case of the S&P index, for instance, with no shortage of significant “market corrections”, one might still accumulate an advantage by stacking up profits at a rate higher than losses. Working with a trading “bias” (or unilaterally) on the buy or sell side could also be explored. For rapidly **decaying** instruments, one may take advantage of mechanisms like those previously mentioned.
For delaying mechanisms, in the long run, the impact of potential orders that may be left “stranded” (or queued for future execution) because of improbable fill prices may cost less than the cumulative PnL improvement over time. This may also entail developing effective algorithms to manage the queue of potential “delayed” orders, enabling the strategy to proceed even with stranded orders. At this point, the question would become essentially whether the gains can outweigh the rate at which drawdowns or losses accumulate.

We introduced here the general concept of using HTI in its most basic form. Future lines of development and research on the concepts presented here could include developing HTI exploitation mechanisms suitable for real-world applications in more complex scenarios. The objective functions in such cases could also **include metrics that account for different manifestations of trading risk**, depending on the portfolio manager's and investors’ goals and mechanisms to limit drawdowns. This could include exploring the distributions of various **reward-to-risk ratios**. For instance, one could consider the expected tail return (Rachev et al., 2004) or various ratios of the PnL to risk measures, such as the maximum or average DD (possibly raised to a suitable power to properly “weight” the importance of the DD), the downside mean square error of the PnL, the current margin requirements (where applicable), or any other meaningful **performance measure**.

These are all research directions that necessitate careful and comprehensive consideration of the individual characteristics of real-world financial instruments. In fact, when managing investors’ funds in real money accounts, we need an in-depth understanding of instruments’ price dynamics as well as sophisticated strategy design to work synergistically in order to effectively **exploit the self-generated HTI**, which is vastly more important than simply “*chasing the price, and finding ways to rationalize it*” (Damodaran, 2021) with some technical indicators.

## Acknowledgments

I am grateful to an anonymous referee of an earlier (unpublished) version of this paper for his thorough and insightful review and encouraging feedback, which led to the current version available on ArXiv.

I am also very grateful to the visionary investors who took a substantial risk on these novel ideas and allowed me to test them in large real-money accounts, thereby encouraging the motivated development and implementation of new hedging mechanisms suitable for dealing with real-world financial instruments.

(Proposals to publish a final, suitably revised version of the current presentation, as well as possible future contributions along this line of research, in suitable peer-reviewed journals, are welcome and will be carefully considered.)